\documentclass[aps,prl,showpacs, amsmath,amssymb, superscriptaddress,lengthcheck]{revtex4-1}
\usepackage{graphicx}
\usepackage{dcolumn}
\usepackage{bm}
\usepackage{color}
\usepackage[T1]{fontenc}
\usepackage[colorlinks,linkcolor=blue,citecolor=blue,urlcolor=blue]{hyperref}
\usepackage{multibib}

\definecolor{darkred}{rgb}{0.85,0,0}

\begin{document}

\title{Thermalization and Cooling of Plasmon-Exciton Polaritons: Towards Quantum Condensation}

\author{S.R.K. Rodriguez} \email{s.rodriguez@amolf.nl}
\affiliation{Center for Nanophotonics, FOM Institute AMOLF, c/o
Philips Research Laboratories, High Tech Campus 4, 5656 AE
Eindhoven, The Netherlands}

\author{J. Feist} \email{johannes.feist@uam.es}
\affiliation{Departamento de F\'isica Te\'orica de la Materia Condensada and Condensed Matter Physics Center (IFIMAC), Universidad Aut\'onoma de Madrid, 28049 Madrid, Spain}
\author{M.A. Verschuuren}
\affiliation{Philips Research Laboratories, High Tech Campus 4, 5656 AE Eindhoven, The Netherlands}

\author{F.J. Garcia Vidal}
\affiliation{Departamento de F\'isica Te\'orica de la Materia Condensada and Condensed Matter Physics Center (IFIMAC), Universidad Aut\'onoma de Madrid, 28049 Madrid, Spain}

\author{J. G\'omez Rivas}
\affiliation{Center for Nanophotonics, FOM Institute AMOLF, c/o
Philips Research Laboratories, High Tech Campus 4, 5656 AE
Eindhoven, The Netherlands}
\affiliation{COBRA Research Institute,
Eindhoven University of Technology, P.O. Box 513, 5600 MB Eindhoven,
The Netherlands}

\date{\today}

\begin{abstract}
We present indications of thermalization and cooling of quasi-particles, a precursor for quantum condensation, in a plasmonic nanoparticle array. We investigate a periodic array of metallic nanorods covered by a polymer layer doped with an organic dye at room temperature. Surface lattice resonances of the array---hybridized plasmonic/photonic modes---couple strongly to excitons in the dye, and bosonic quasi-particles which we call plasmon-exciton-polaritons (PEPs) are formed. By increasing the PEP density through optical pumping,  we observe thermalization and cooling of the strongly coupled PEP band in the light emission dispersion diagram. For increased pumping, we observe saturation of the strong coupling and emission in a new weakly coupled band, which again shows signatures of thermalization and cooling.
\end{abstract}
\pacs{73.20.Mf, 33.57.+c,  71.36.+c, 42.50.-p}

\maketitle

Surface plasmon polaritons (SPPs) are light-matter quasi-particles at a metal-dielectric interface, enabling control of light on subwavelength scales~\cite{Barnes03}. SPPs are bosons, and by virtue of bosonic stimulation, transition rates into a quantum state are enhanced when the final state occupation exceeds unity. Bosonic stimulation underlies the laser through stimulated emission, and condensation through stimulated scattering. The former has allowed plasmonics to open a new era of nanoscopic coherent light sources~\cite{Stockman03,Noginov09,Oulton09,Hess12}. In contrast, condensation of SPPs into a single quantum state appears to have never been considered. The reasons for this are likely manifold. Propagating SPPs do not have a cut-off---their ground state is at zero frequency, such that thermalization is not number-conserving and condensation does not occur. On the other hand, localized surface plasmon resonances (LSPRs) have a flat dispersion, implying infinite effective mass. Condensation is more easily achieved with low-mass quasi-particles, as it occurs when the mean thermal wavelength $\lambda_T\!\propto\!(mk_BT)^{-1/2}$ exceeds the interparticle spacing. Additionally, the quasi-particles have to thermalize, which poses a challenge for plasmonic systems with typical lifetimes $\lesssim 10$ fs.

One system that may overcome the aforementioned limitations is a periodic array of metallic nanoparticles covered by organic molecules in solid-state. LSPRs in the nanoparticles hybridize with diffraction orders radiating in the plane of the array (so-called Rayleigh anomalies), leading to surface lattice resonances (SLRs)~\cite{Zou&Schatz04b,Garcia07, Crozier08, Auguie&Barnes08, Kravets08, Vecchi09}.  While the SPP-exciton strong coupling has been investigated for propagating modes in flat~\cite{Bellessa2004, Torma09, Gonzalez-Tudela2013} and  perforated~\cite{Dintinger2005, Vasa2008, Schwartz11} metallic layers, as well as for localized modes in nanostructures~\cite{Sugawara06, Cade09, Manjavacas}, the strong coupling of SLRs to excitons remains unexplored. Advantageously, SLRs can have a narrow linewidth (few meVs~\cite{Zou&Schatz04b}) and tunable dispersion via the nanoparticle geometry and lattice constant~\cite{Odom11, Rodriguez11}, thereby supporting low-mass polaritons with relatively long lifetimes. As shown ahead, strongly coupled SLR-exciton quasi-particles are low-mass ($\sim 10^{-7}$ times the electron rest mass) analogues of exciton-polaritons in semiconductor microcavities, for which condensation has been observed in several groundbreaking experiments~\cite{Deng02,Kasprzak06,Balili07}. We therefore call them plasmon-exciton-polaritons (PEPs).

In this Letter, we demonstrate the suitability of PEPs in metallic nanoparticle arrays for quantum condensation. We show that they thermalize, with their effective temperature approaching the lattice temperature when their density is increased through optical pumping. In the present system, we observe a saturation of the strong SLR-exciton coupling before condensation sets in. This leads to a transition from  strong to weak coupling, after which we observe thermalization and cooling of the weakly coupled SLR mode.

\begin{figure}
\includegraphics[width=\linewidth]{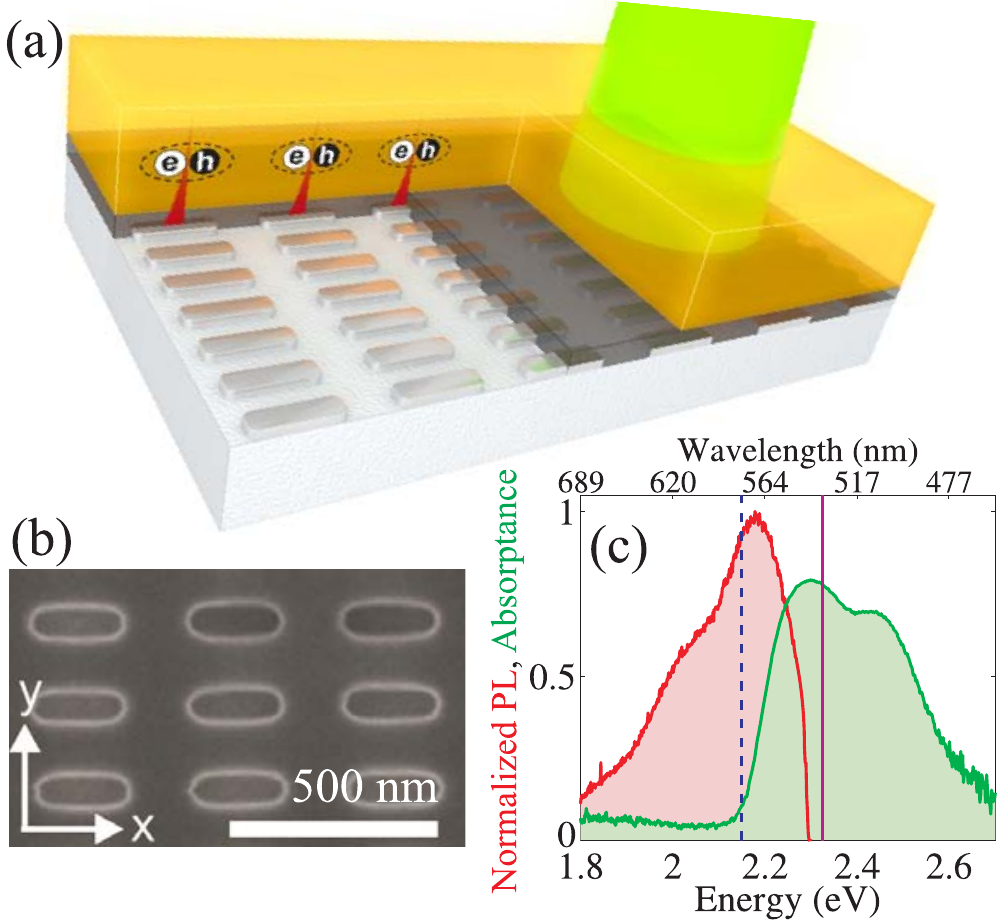}
\caption{(a) A silver nanorod array on an SiO$_2$ substrate covered by a thin Si$_3$N$_4$ layer (gray) and a R6G/PVA layer (orange). An incident laser (green) pumps the R6G exciton reservoir. (b) Scanning electron microscope image of the resist layer used for the fabrication of the nanorod array.  (c) Normalized photoluminescence (red) and absorptance (green) of the R6G layer without the nanorod array. The solid line indicates the pump energy, while the dashed line indicates the emission energy of the saturated ground state at the highest pump power.} \label{fig1}
\end{figure}	

Figure~\ref{fig1} illustrates the sample. A periodic array of silver nanorods was fabricated onto a fused silica substrate by substrate conformal nanoimprint lithography~\cite{SCIL}. A scanning electron microscope image of the resist layer used for the fabrication is shown in \autoref{fig1}(b). The rod dimensions are $230 \times 70 \times 20$ nm$^3$, with lattice constants $a_x = 380$ nm and $a_y=200$ nm.  A $20$ nm layer of Si$_3$N$_4$ on top of the array prevents the silver from oxidizing. A $300$ nm layer of polyvinyl alcohol (PVA)---with Rhodamine 6G (R6G) dye molecules for the emission experiments---was spin-coated on top. Figure~\ref{fig1}(c) shows the absorptance and the normalized emission of the R6G layer. All experiments were performed at room temperature ($300$ K). Further details are included in the supplemental material~\cite{supplemental}.

\begin{figure}[tb]
\includegraphics[width=\linewidth]{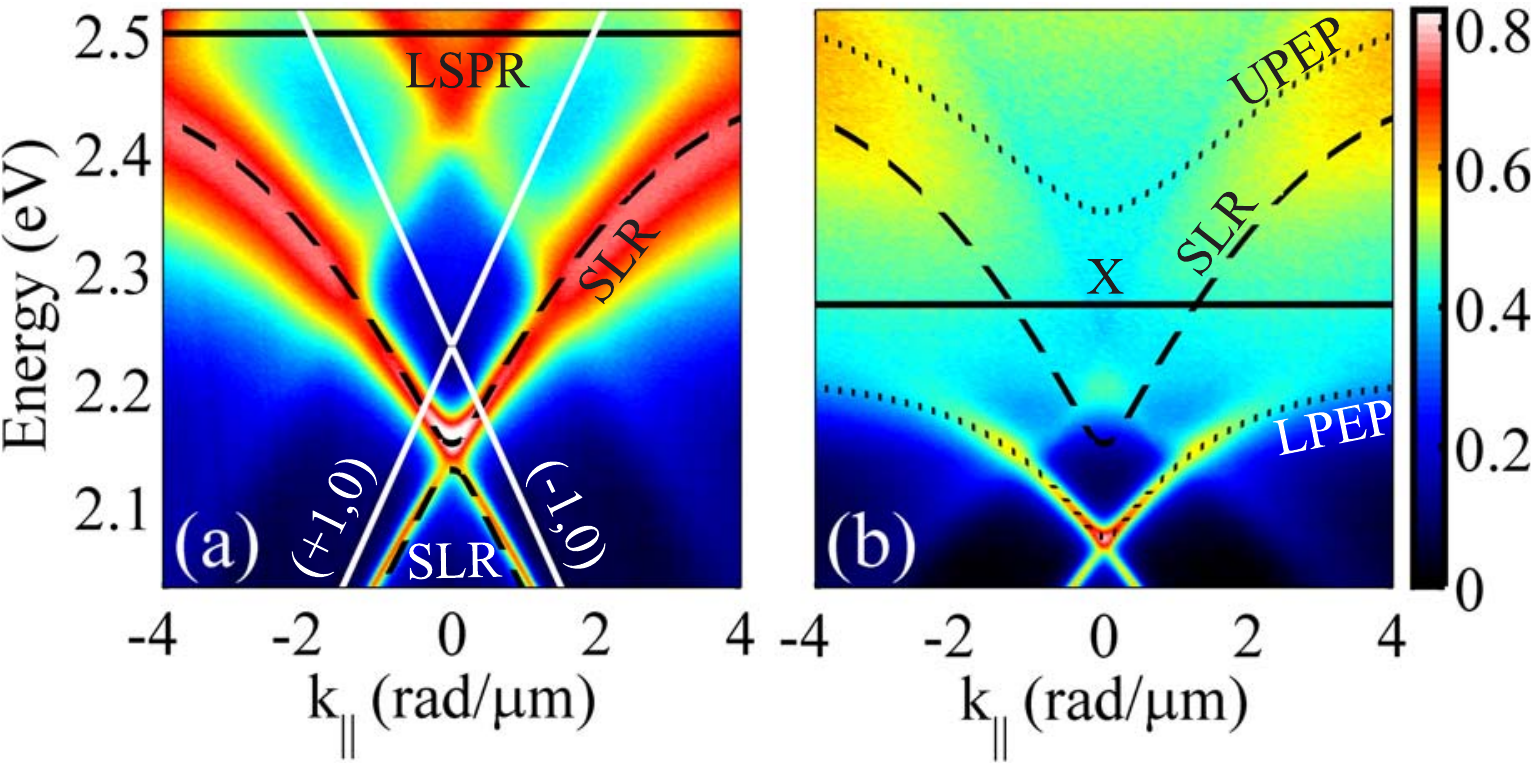}
\caption{Extinction spectra of the nanorod array covered by a polymer layer (a) without  and (b) with R6G molecules, both in the same color scale. In (a): the solid white lines indicate the Rayleigh anomalies, while the solid black line indicates the localized surface plasmon resonance. The  dashed lines indicate the surface lattice resonances. In (b): The solid line indicates the R6G exciton, the dashed line is the upper SLR from (a), and the  dotted lines indicate the mixed states (plasmon-exciton-polaritons).}\label{fig2}
\end{figure}

We first analyze the strongly coupled modes in the nanorod array. Figure~\ref{fig2} shows two light extinction measurements: In \autoref{fig2}(a) the PVA layer has no R6G molecules, while in \autoref{fig2}(b) R6G molecules were embedded at 23 weight $\%$ with respect to the PVA (R6G number density $\approx 3.6\cdot10^8\,\mu$m$^{-3}$). The extinction, defined as $1-T_0$ with $T_0$ the zeroth-order transmittance, is shown in color as a function of the incident photon energy and  wave vector  component $k_\|$ parallel to the long axis of the nanorods (x-axis).  The incident light is $s$-polarized, probing the short axis of the nanorods (y-axis).

The dispersive extinction peaks in \autoref{fig2}(a) are SLRs associated with the $(\pm1,0)$ diffraction orders. These hybrid modes arise from the coupling between the LSPR (solid black line) and the ($\pm1,0$) Rayleigh anomalies (intersecting solid white lines). The SLRs (dashed black lines) are calculated with a $3\times3$ model Hamiltonian including the decay of the three modes and their mutual coupling (see supplemental material for details~\cite{supplemental}). A small gap between the upper and lower SLRs is observed at normal incidence near $2.15$ eV, where only the upper SLR is excited due to the mode symmetries. The electric field distribution in the plane of the array is symmetric for the upper SLR, but antisymmetric for the lower SLR, which renders the former ``bright'' and the latter ``dark''~\cite{Rodriguez11}.

Figure~\ref{fig2}(b) shows the extinction of the same array but with the R6G molecules embedded in the PVA. The solid line indicates the peak energy of the R6G exciton. The dashed line is the upper SLR as shown in \autoref{fig2}(a). PEPs are composite quasi-particles emerging from the strong coupling of these two resonances, which we model with a $2\times2$ Hamiltonian~\cite{supplemental}. The inhomogeneity in the coupling between molecules and SLRs enters into the properties of the collective Dicke state that forms the excitonic part of the $2\times2$ Hamiltonian, but does not further influence the dynamics~\cite{Houdre1996,Gonzalez-Tudela2013}. We ignore the LSPR and lower SLR because they have minimal influence in what follows. The upper and lower PEPs, indicated by the dotted lines in \autoref{fig2}(b), display a $250$ meV Rabi splitting at zero SLR-exciton detuning. The extinction of the upper PEP band is smeared out due to the increased SLR linewidth at higher energies, and possibly due to the influence of another mode [see near $2.4$ eV at $k_\|=0$ in \autoref{fig2}(a)]. However, this has minimal influence on the lower PEP. Analogous to the Hopfield coefficients~\cite{Hopfield},  the exciton coefficient $x(k_\|)$ and SLR coefficient $s(k_\|)$ characterize the relative contributions to the PEP.  We find $|x(0)|^2 = 0.3$ and $|s(0)|^2 = 0.7$ for the lower PEP at $k_\|=0$. Thus, despite the large SLR-exciton detuning ($-118$ meV), the exciton fraction is not negligible. Near $k_\|=0$, the effective mass of the lower PEP is $m^{\ast}_{p} \approx \hbar^2 / ( \partial^2 E / \partial k_\|^2) \approx 2.0 \cdot 10^{-37}$ kg. These PEPs are $10^{10} - 10^{12}$ times lighter than atoms~\cite{Anderson95, Ketterle95}, and $\sim\!100$ times lighter than exciton-polaritons~\cite{DengRMP}. The characteristic temperatures ($\sim\!1000\,$K) in our system are correspondingly higher (keeping the mass-to-temperature ratio similar): about $10^8 - 10^{11}$ times higher than in atomic Bose-Einstein condensation systems~\cite{Anderson95, Ketterle95}, and $\sim\!10^2$ times higher than GaAs and CdTe exciton-polariton systems~\cite{DengRMP}. Note that recent GaN and ZnO exciton-polariton systems can operate at room temperature~\cite{Lai13}.

\begin{figure}[tb]
\includegraphics[width=\linewidth]{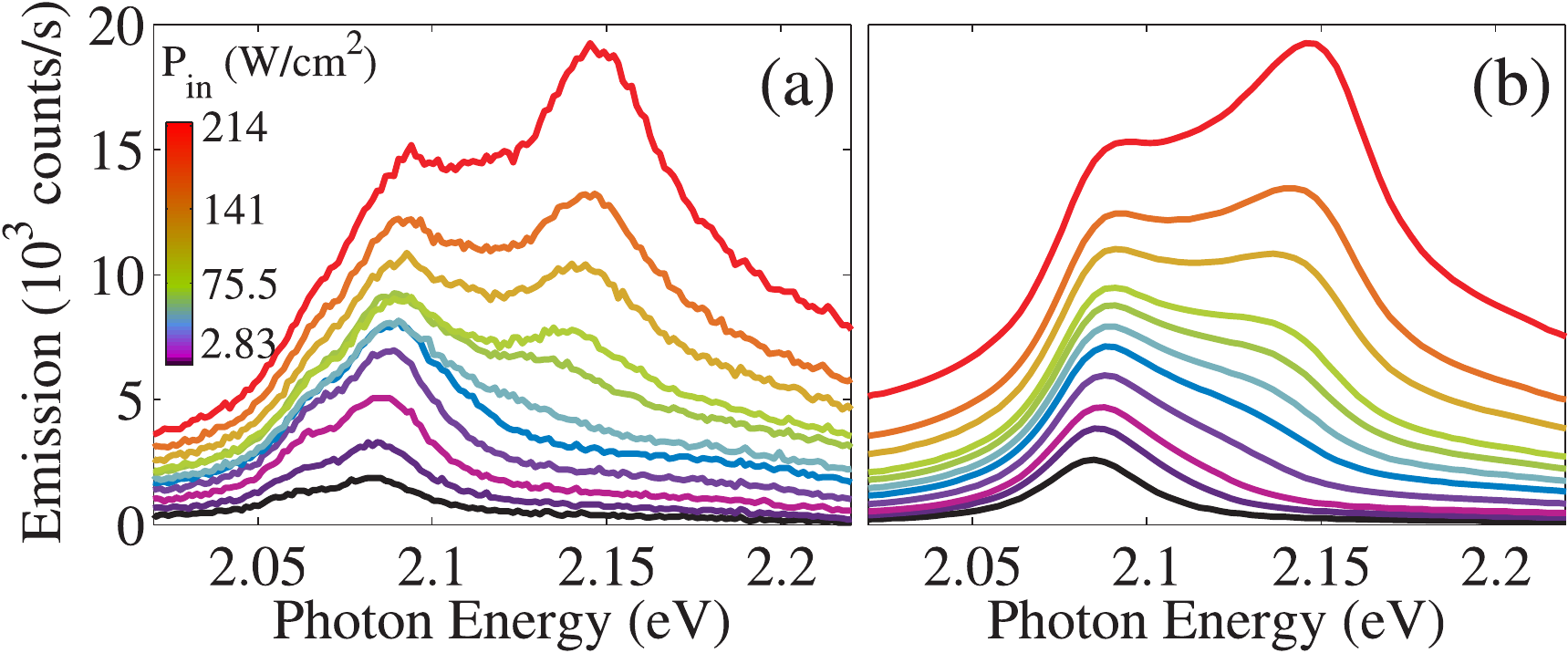}
\caption{ Emission at $k_\|=0$, (a) for different experimental input power densities encoded in color, and (b) predicted by a model based on local saturation of the coupling (see text for details).}
\label{fig3}
\end{figure}

In \autoref{fig3}  we present a series of emission measurements obtained by increasingly pumping the structure in \autoref{fig2}(b) with an optical parametric oscillator having peak energy $2.33$ eV, $\sim\!200$ fs pulses, and $80$ MHz repetition rate. The sample is fixed, while the detector rotates collecting s-polarized light with in-plane momentum $k_\|$ along $\hat{x}$~\cite{supplemental}. \autoref{fig3}(a) shows the forward ($k_\|=0$) emission spectrum as a function of the pump irradiance.  The peak at $\sim\!2.08$ eV, which dominates the spectrum below a critical irradiance of $P_c \approx 60$ W/cm$^2$, is the emission from the lower PEP. The shoulder at 2.065 eV  is attributed to the lower SLR, which is dark at $k_\|=0$ but appears in the spectrum due to the finite angular resolution of the experiment. As pumping increases, the lower PEP peak blue-shifts and broadens. Above $P_c$ a new peak emerges at $\sim\!2.15$ eV, and at $\sim\!2 P_c$ its emission exceeds the lower PEP emission. We attribute the shift of the coupled states towards the uncoupled states to saturation of the coupling with increasing exciton density~\cite{Houdre95,Butte02}. This has previously been observed in plasmon-exciton polariton systems as a diminished normal mode splitting in the frequency-domain, and as a reduced Rabi frequency in the time-domain~\cite{Vasa10,Vasa13}.

\begin{figure*}[tb]
\includegraphics[width=\linewidth]{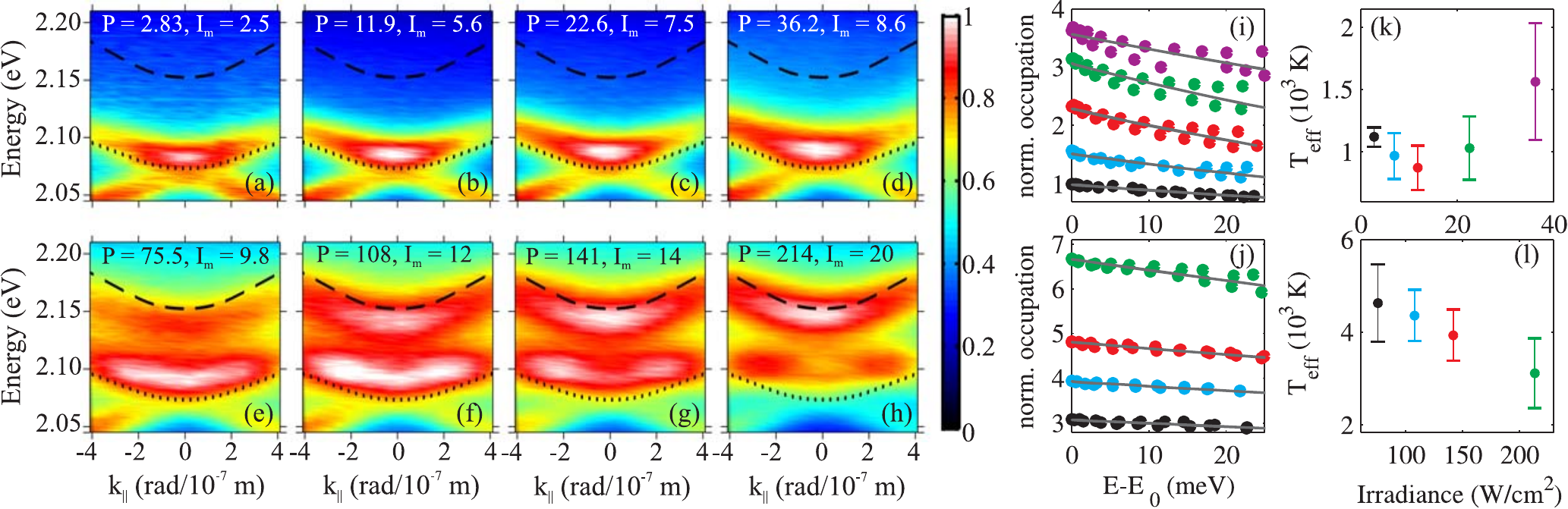}
\caption{(a-h) Emission for increasing pump irradiance $P$ (labeled in W/cm$^2$). The colorbar pertains to all spectra, with the maximum intensity $I_{\text{m}}$ indicated in $10^3$ counts/s. The dashed and dotted lines are the SLR and PEP energies as in \autoref{fig2}(b). The scale is different from \autoref{fig2} to magnify the emission near $k_\|=0$. (i,j) Normalized occupation of the (i) lower (PEP) band below criticality and (j) upper (SLR) band above criticality. Data points of different colors correspond to different $P$. The two groups of points for each $P$ correspond to the occupation for $\pm k_\|$, i.e. the data is slightly asymmetric. The solid gray lines are Maxwell-Boltzmann fits. The mapping between $P$ and color can be inferred from (k) and (l), which show effective temperatures retrieved from the fits in (i) and (j), respectively.}
\label{fig4}
\end{figure*}

Instead of a smooth transition of the peak energy, we observe the coexistence of two bands at intermediate pump powers. This is explained by a model taking into account the spatiotemporal profile of the excitation density. We assume that the exciton-SLR coupling saturates as $\Omega_{\text{XS}}=\Omega_{\text{XS},0}(1+n(r,t)/n_{\text{sat}})^{-1/2}$, with the excitation density $n$ varying spatially over the pump beam profile and decaying in time after the pump. The emission from each point in space and time is then assumed to be at the PEP energy given by inserting this $\Omega_{\text{XS}}$ into the two-state model used for modeling the PEP. At zero detuning, strong coupling occurs when the energy exchange rate $\Omega_{\text{XS}}$ is larger than the decay rates $\gamma_X$ and $\gamma_S$ of the exciton and SLR, respectively. Although the distinction becomes somewhat ambiguous for non-zero detuning as in the present case, the two extreme cases can be readily identified: Small $n$ leads to strong coupling ($\Omega_{XS}\gg \gamma_X,\gamma_S$), while $n\gg n_{\text{sat}}$ gives weak coupling ($\Omega_{XS} \ll \gamma_X,\gamma_S$). Integrating over space and time gives the model spectra shown in \autoref{fig3}(b)~\cite{supplemental}, which agree well with the experimental data. From the model, it follows that the low-energy peak stems from regions with low $n$ where the emission from the strongly coupled PEPs dominates, while the high-energy peak stems from regions with high $n$ where the emission from the weakly coupled SLR dominates. The intermediate regions form a broad background that is not resolved as an isolated peak. Accordingly, for increased pumping the lower PEP emission saturates, while the new band blue-shifts towards the bare SLR state and grows in intensity. The experimental output vs.\ input power dependence confirms this~\cite{supplemental}.

We now study the dispersion of the observed bands through their angle-dependent emission intensity, shown in Figures~\ref{fig4}(a)-(h) for several pump powers. We find that the lower PEP band is slightly blue-shifted compared to the extinction measurements. It is also flatter, corresponding to a slightly higher effective mass of $\approx 2.6 \cdot 10^{-37}\,$kg (extracted from the curvature of the band at $k_\|=0$). We attribute this to a well-known difference between extinction and emission spectra: Extinction stems from the interference between direct and scattered radiation, while emission does not contain a direct part. This leads to a shift of the peak emission energy~\cite{Rodriguez12PRL}, which can also affect the extracted effective mass if it is angle-dependent. For high pumping, the peak of the emission is slightly red-shifted from the SLR peak in extinction. According to our model, this implies that the coupling is not fully saturated---the underlying SLR energy used for the model in \autoref{fig3}(b) is again slightly blue-shifted compared to extinction~\cite{supplemental}.

Next, we discuss the thermalization behavior. Condensation as a thermodynamic phase transition requires the system to approach thermal equilibrium, which constrains the ratio of thermalization to decay time. For inorganic exciton-polaritons, both of these times are $1-10\,$ps, and consequently both equilibrium and non-equlibrium condensation have been observed~\cite{DengRMP}. For the present system, we estimate a PEP lifetime of at least $\sim\!17\,$fs from the emission linewidth. Vibrational relaxation of R6G, and thus PEP-phonon scattering, occurs on a scale of $\sim\!100\,$fs~\cite{Elsa}, while PEP-PEP scattering rates are currently unknown. Therefore, equilibrium dynamics seem unlikely in our case. As we show next, we nevertheless observe thermalization and cooling for increased pumping, possibly due to more efficient PEP-PEP scattering at high density.

Figures~\ref{fig4}(a)-(d) display a greater emission from the strongly coupled band at low pumping, while Figs.~\ref{fig4}(e)-(h) display a greater emission from the weakly coupled band at high pumping.  We study this in detail by analyzing the occupation $n_{oc}$ as a function of the emitted photon energy, shown in \autoref{fig4}(i) for the strongly coupled band and in \autoref{fig4}(j) for the weakly coupled band. The occupation is extracted from the emission intensity $I(k_\|)$ along the corresponding band, integrated over a fixed bandwidth of $40$ meV. We take into account that PEPs are composite quasi-particles and only their photonic component leaks out of the open system. Thus, as in exciton-polariton systems~\cite{Tartakovskii2000, Butte02}, we correct for the SLR fraction $|s(k_\|)|^2$, giving $n_{oc} \propto I(k_\|) / |s(k_\|)|^2$. Here we have assumed that SLRs mainly decay radiatively due to their large Rayleigh anomaly fraction and the predominantly radiative decay of LSPRs. The gray lines in \autoref{fig4}(i) and 4(j) are fits of the occupation to a Maxwell-Boltzmann distribution $n_{oc}\propto \exp[-(E-E_0)/k_B \mathrm{T_{eff}}]$, from which we extract the effective temperature $\mathrm{T_{eff}}$. This is shown as a function of the pump irradiance in \autoref{fig4}(k) and \autoref{fig4}(l) for the strongly and weakly coupled band, respectively. The error bars represent a $2\sigma$ ($\approx95\%$) confidence interval, and stem mostly from a small asymmetry of $n_{oc}$ for positive and negative $k_\|$. This is possibly due to angle-dependent variations in collection efficiency and intensity fluctuations during the measurement time.

The effective temperature of the strongly coupled PEPs displays an initial decrease, but remains warmer than the lattice. This observation, which is reminiscent of early works on exciton-polariton condensates~\cite{Deng02, Balili07}, indicates that the system approaches but does not fully reach thermal equilibrium with the heat bath (the molecule phonons). In addition, we observe that for increased pumping the ground state ($E-E_0=0$) occupation increases slightly above the Maxwell-Boltzmann fit. This could be an indication that the bosonic statistics of the PEPs are becoming relevant, implying that condensation is being approached, although not reached. Consequently, the ground state occupation remains much lower than in  exciton-polariton condensates~\cite{Kasprzak06,Balili07}. As the power increases and saturation is approached, $\mathrm{T_{eff}}$ increases again, although the experimental uncertainty from the fits also increases. This increased uncertainty implies a stronger deviation of the $n_{oc}$ from a thermal distribution near the strong-to-weak coupling transition. Therefore, while cooling of PEPs is observed, saturation of the SLR-exciton coupling sets in before condensation is reached, and the new band with weaker coupling emerges and blue-shifts towards the bare SLR state. The effective temperatures in this new band [\autoref{fig4}(l)] are higher than in the PEP band. Nevertheless, $\mathrm{T_{eff}}$ decreases monotonically as pumping increases. This cooling implies that condensation of bare SLRs could be within reach, analogous to the condensation of cavity photons observed by Klaers \emph{et~al.}~\cite{Klaers2010}. Currently, further pumping was not possible because the molecules bleached. A different pump source or dye could circumvent this limitation, enabling higher excitation densities. We note that SLR lasing has recently been demonstrated by Zhou and co-workers~\cite{Odom13}.

In conclusion, we presented experimental indications of thermalization and cooling of quasi-particles in an array of Ag nanoparticles covered by organic molecules. This array supports surface lattice resonances, which form plasmon-exciton-polaritons (PEPs) through strong coupling to molecular excitons.  In view of the low PEP mass, which is furthermore tunable via the surface lattice resonance dispersion, we believe that plasmonics holds great promise for solid-state studies of macroscopic quantum many-body physics at and above room-temperature. While the short lifetimes of plasmons make thermodynamic equilibrium challenging, we envisage these results to open a new avenue for studying non-equilibrium quantum dynamics.

\acknowledgments
We thank Francesca Marchetti for stimulating discussions. This work was supported by the Netherlands Foundation for Fundamental Research on Matter (FOM) and the Netherlands Organization for Scientific Research (NWO), and is part of an industrial partnership program between Philips and FOM. J.F. and F.J.G.V. acknowledge support by the European Research Council under Grant No. 290981 (PLASMONANOQUANTA).

\clearpage
\section{Supplemental Material}
\subsection{Experimental details}
The extinction and photoluminescence (PL) measurements presented in the main text were all performed with computer-controlled rotation stages. The angular resolution was $0.25^{\circ}$ degrees for all measurements. For the extinction, we measured the transmittance through the array of a collimated (angular spread $< 0.1^{\circ}$) and linearly polarized light beam from a halogen lamp. The spot size was $500~\mu$m. We define the extinction  as $1-T_0$, with $T_0$ the zeroth-order transmittance. Reflectance measurements are not included, but can be found for similar nanoparticle arrays in Ref.~\cite{Vecchi09b}. For the PL measurements, the pump was an optical parametric oscillator with peak energy $2.33$ eV, $\sim\!200$ fs pulses, and $80$ MHz repetition rate. The pump beam was lightly focused (spot size of $130~\mu$m) and impinged with $k_\|=2.05$ rad/$\mu$m. The emitted light was collected by a fiber-coupled spectrometer preceded by a polarization analyzer.  The pump power was varied with a neutral density filter, and measured in front of the sample. The highest power used yielded a fraction of excited R6G molecules below $10^{-3}$, as discussed ahead. Both extinction and PL measurements are shown for s-polarized light (y-axis, parallel to the short axis of the nanorods) with in-plane momentum $k_\|\hat{x}$ (parallel to the long axis of the nanorods). All experiments were performed at room temperature ($300$ K).

An important experimental parameter in the PL measurements is the time duration of the pump pulse. The pulse duration was roughly a factor of 2 higher than the expected vibrational relaxation time scale of the organic molecules, which poses a challenge for an equilibrium state to be established.  The unlikelihood of an equilibrium state with a true temperature in the usual sense is already manifest in the high effective temperatures reported in Figs.\ 4(k) and 4(l) in the main text. The fact that these effective temperatures are much higher than the lattice temperature indicates that while plasmon-exciton-polaritons may achieve local self-equilibrium amongst themselves through inelastic scattering, they remain out of equilibrium with their surroundings. However, the relatively short pulse duration was chosen for a reason: A shorter pulse yields higher peak powers compared to continuous-wave or long-pulse lasers with equal average power. The high peak power leads to a high density of excitations, while the removal of the pump power shortly after (at the end of the pulse) assists in heat dissipation. Thus, greater ``instantaneous'' densities---which are essential for the results in this Letter---can be created before reaching the damage threshold of the molecules. In fact, this damage threshold (an irreversible photobleaching) prevented us from using higher pump powers in the PL experiments.
As the system dynamics depend sensitively on the excitation source, we expect future experiments using different sources to reveal novel phenomena, and to further clarify the influence of the pulse duration on the thermalization dynamics.

\subsection{Modeling surface lattice resonances and plasmon-exciton-polaritons}
Here we describe the few-state models employed in deriving the surface lattice resonance (SLR) and plasmon-exciton-polariton (PEP) dispersion relations in Fig.~2 in the main text.
We calculate the SLRs [Fig.~2(a) in the main text] with a $3\times3$ complex Hamiltonian,
\begin{equation}\label{eq1}
H = \left( \begin{matrix}
E_{L} - i \gamma_L   & \Omega_{L+}          & \Omega_{L-}  \\
\Omega_{L+}         &  E_{R+}  - i \gamma_{R+} & \Omega_{\pm}    \\   
 \Omega_{L-}        &  \Omega_{\pm}            & E_{R-}  - i \gamma_{R-} \end{matrix} \right),
\end{equation}
where the subscripts ``$L$'', ``$R+$'', and ``$R-$'' stand for LSPR, (+1,0), and (-1,0) Rayleigh anomaly, respectively.
We set $E_{L} - i \gamma_L = (2.5 - 0.12i)\,$eV for the LSPR, and $\gamma_{R+} = \gamma_{R-} = 10\,$meV for the Rayleigh anomaly losses.  The dispersion relation $E_{R\pm}(k_\|)$ of the ($\pm 1,0$) Rayleigh anomalies follows from the conservation of the parallel component of the wave-vector: $E_{R\pm}(k_\|) = \frac{\hbar c}{n_{\mathrm{op}}}  |  k_\|  + m G_x |$, where $m$ is the order of diffraction, $G_x = \frac{2 \pi}{a_x} $ is the $x$-component of the reciprocal lattice vector, and $n_{\mathrm{op}}=1.46$ is the refractive index. $E_{R\pm}(k_\|)$ are plotted in Fig.~2(a) in the main text as solid white lines. Diagonalization of this Hamiltonian yields the complex SLR energies, with real parts plotted as dashed lines in Fig.~2(a). The coupling strengths $\Omega_{L+}  = \Omega_{L-} = 184\,$meV and $\Omega_{\pm} = 105\,$meV are fitted to reproduce the experimentally observed dispersion.

The plasmon-exciton-polariton (PEP) dispersion [Fig.~2(b) in the main text] is obtained similarly. The relevant states are now the upper SLR and the R6G exciton. PEPs emerge from the strong coupling of these two resonances. As the lower SLR and the highest state ($\sim$LSPR) have minimal influence on the dynamics, we neglect them in the following. The PEP energies are thus calculated with a $2\times2$ Hamiltonian similar to \autoref{eq1}. The SLR energies are obtained from the diagonalization of \autoref{eq1}, while for the exciton we set $E_{\text{X}} - i \gamma_{\text{X}} = (2.27- 0.1i)\,$eV. The mutual SLR-exciton coupling $\Omega_{\text{XS}} = 0.127\,$eV is again obtained by fitting to the experimental results. The eigenenergies of this model Hamiltonian (the upper and lower PEP bands) are indicated by the dotted lines in Fig.~2(b) in the main text.

\subsection{Excited molecule density}
Here we estimate the fraction of excited R6G molecules by each laser pulse. The pump laser has a photon energy $E_{\text{pump}}\approx2.33\,$eV and a repetition rate $f=80\,$MHz, corresponding to a pulse every $12.5\,$ns. As the longest lifetime of any excitation in the system (in the ps range) is orders of magnitude shorter than the repetition rate, all excitations decay before the next pulse arrives. For the highest average pump irradiance, $P_{\text{max}}=214\,$W/cm$^2$, the 2D photon density per pump pulse is $\rho_{\text{pump}} = P_{\text{max}}/(f E_{\text{pump}}) \approx 7.25\cdot10^4\,\mu$m$^{-2}$.

The R6G molecules are embedded in the PVA layer at 23 weight \%, corresponding to a density of $0.29\,$g/cm$^3$, and a number density of $n_{\text{R6G}}\approx 3.6\cdot10^8\,\mu$m$^{-3}$. The thickness of the layer is $t_{\text{PVA}}= 300\,$nm, giving a 2D density of $n_{\text{R6G}} t_{\text{PVA}} \approx 1.1\cdot10^8\,\mu$m$^{-2}$.
The absorption cross section of R6G at $E_{\text{pump}}$ and at high concentration is on the order of $\sigma\approx 2\cdot10^{-8}\,\mu$m$^2$~\cite{Lu1986}.
This gives an optical density of $\mathit{OD} = \sigma n_{\text{R6G}} t_{\text{PVA}} \approx 2.2$, and a corresponding absorptance of $A = 1-\exp(-\mathit{OD})\approx 0.89$. This agrees reasonably well with the measured absorptance of the bare layer [Fig.~1(c) in the main paper]. Thus, almost all photons of the incoming pulse are absorbed. The 2D excitation density is then $A \rho_{\text{pump}} \approx 6.5\cdot10^4\,\mu$m$^{-2}$, which means that the fraction of excited molecules is about $5.9\cdot 10^{-4}$.

\subsection{Output vs. input power dependence}
\begin{figure}
\includegraphics[width=\linewidth]{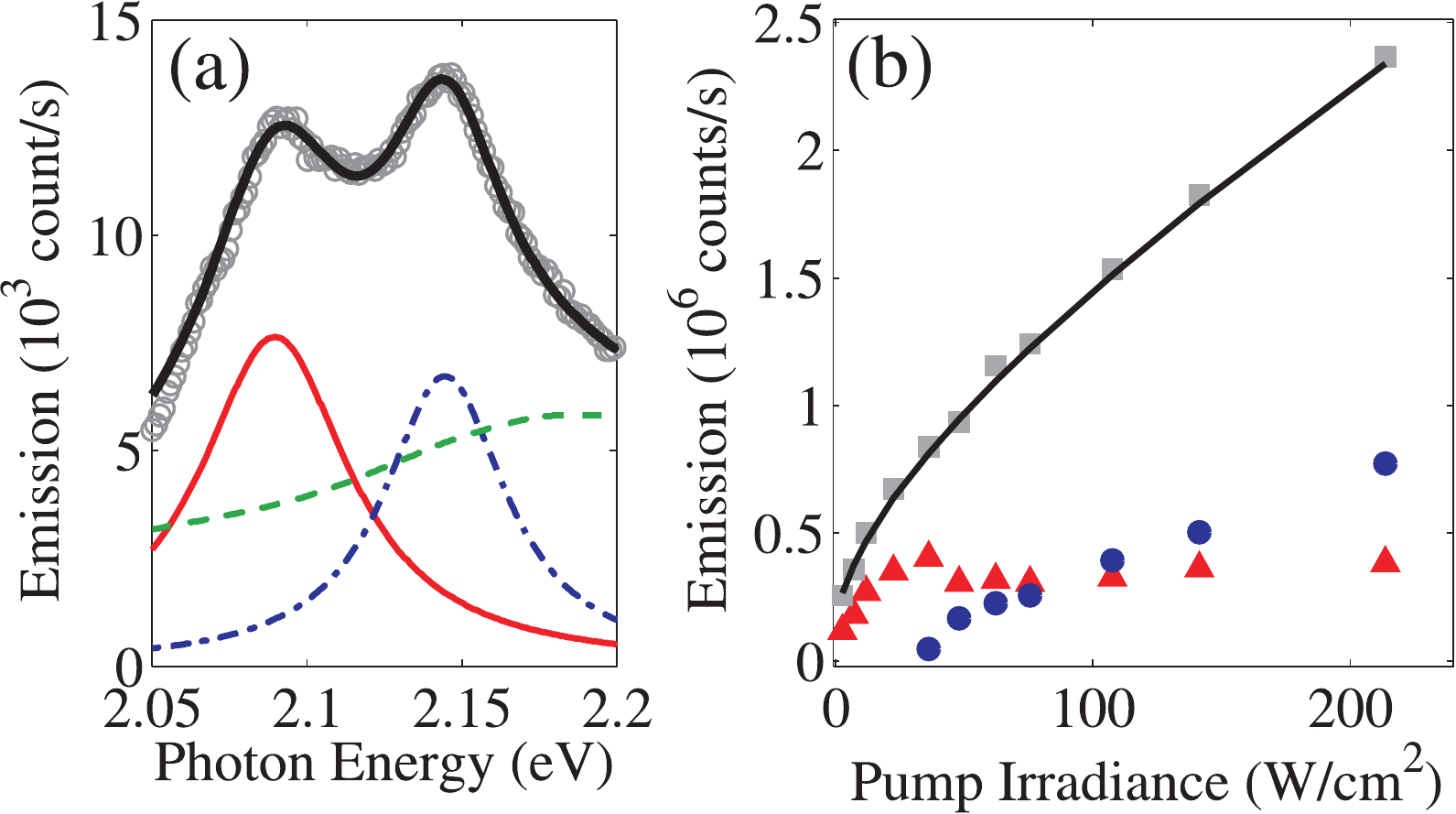}
\caption{ (a) The grey circles are the measured emission intensity at $k_\|=0$ for $P=141$ W/cm$^2$. This is fitted (black solid line) with the sum of two Lorentzians (red solid line and blue dash-dotted line) and the background R6G emission (green dashed line). (b) Emission for the different states as a function of the pump irradiance. The red triangles correspond to the lower plasmon-exciton-polariton which saturates. The blue circles correspond to the upper band approaching the bare SLR state. The grey squares are the total output power obtained by combining the previous two and the background R6G emission. The black line is a fit to the experimental data, with linear and square root terms as described in the text.} \label{figS1}
\end{figure}

Here we analyze the measured emission intensity at $k_\|=0$ to extract the contributions from the strongly coupled PEPs and weakly coupled SLRs at different pump powers. To this end, we fit the measurements with the sum of two Lorentzians and the background R6G emission. As described in detail in the following section, the two peaks observed in the spectra correspond to the extrema of a continuous distribution of emission lines with different SLR-exciton coupling strength. Their lineshapes are thus not expected to be purely Lorentzian, but include (non-Gaussian) inhomogeneous broadening. For simplicity, we use Lorentzian lineshapes in the following. One such fit is illustrated in \autoref{figS1}(a) for $P=141$ W/cm$^2$. Therein, the grey circles are the measured intensity, while the black solid line is obtained by summing the two Lorentzians (red solid line and blue dash-dotted line) and the background R6G emission (green dashed line). The good agreement between the data and the fit partially validates the choice of pure Lorentzians. Similar fits were performed for the different pump powers. From the area under the fitted Lorentzians, we calculated the emitted power by each of the bands as a function of the pump irradiance. This is shown in \autoref{figS1}(b), where the blue circles correspond to the high-energy (SLR-like) peak, the red triangles correspond to the low-energy (PEP-like) peak, and the grey squares are the total emitted power. The black solid line is a fit of the total emitted intensity $I_{out}$ to a function of the form $I_{out} \propto I_{in} + \alpha \sqrt{I_{in}}$, with $I_{in}$ the pump irradiance and $\alpha$ a fit parameter.  The sublinear square-root term is likely due to bimolecular quenching~\cite{Kena-Cohen2010}, and it is taken into account in our saturation model as described next.

\subsection{Saturation model}
Here we describe the model for the emission spectrum under saturation of the SLR-exciton coupling in more detail.
We assume that the energy of the emitting (PEP) mode is given by the (lower) eigenstate of the $2\times2$ Hamiltonian from which we obtain the PEP dispersion as described above.
However, instead of being constant, the exciton-SLR coupling $\Omega_{\text{XS}}$ saturates as a function of the (spatially and temporally) varying excitation density,
$\Omega_{\text{XS}}(n) = \Omega_{\text{XS},0} (1+ n/n_{\text{sat}})^{-1/2}$, where $\Omega_{\text{XS},0}$ is the unsaturated coupling strength, and $n_{\text{sat}}$ is the saturation density.
The local excitation density is taken proportional to the spatially varying pump beam power (as the PEP propagation length is much smaller than the spot size). In time, it follows the decay of the (uncoupled) R6G molecules that are pumped (as the PEP lifetime of $\sim\!17\,$fs is a lot shorter than the picosecond lifetime of the excitations). Assuming a Gaussian beam, this gives $n(r,t) \propto P_{\text{max}} e^{-r^2/r_0^2 -t/t_0}$, where $r=\sqrt{x^2+y^2}$ is the distance from the beam center in the plane, and $r_0$ and $t_0$ are the spot size and excitation lifetime.
The emission from each point follows a Lorentzian energy distribution centered around the (local) PEP energy. The emission intensity is proportional to $n(r,t)+\alpha \sqrt{n(r,t)}$, where the sublinear square-root term is likely due to bimolecular quenching~\cite{Kena-Cohen2010}, and is also observed in the experimental results for the total output power as a function of the pump power (\autoref{figS1}). Therefore, the output power is

\begin{equation}\label{Pout}
P_{\text{out}}(E,r,t) \propto \frac{\left(n(r,t)+\alpha\sqrt{n(r,t)}\right) \gamma} {(E-E_{plex}[n(r,t)])^2 + \gamma^2},
\end{equation}
where $\gamma$ is the emission linewidth, which in principle could also depend on the local density. As the change of $\gamma$ when going from the fully coupled PEP to the bare SLR is expected to be small, it is chosen constant here for simplicity. The total emission spectrum is then given by an integral of \autoref{Pout} over space and time:
\begin{equation}\label{Iout}
I_{\text{out}}(E) \propto \int\limits_{0}^{\infty} \int\limits_{0}^{\infty} 2\pi r P_{\text{out}}(E,r,t) \,\mathrm{d}t\,\mathrm{d}r\,,
\end{equation}
where we have exploited the radial symmetry of the integrand. Note that the final result does not depend on $r_0$ or $t_0$ apart from a global scaling.
In the following, we do not fix the proportionality factor between input power and $n(r,t)$, but instead set $n_{\text{sat}}$ and $\alpha$ in units where $n(r,t)$ is given in W/cm$^2$, and set $P_{\text{max}}$ to the total input power. The parameter values used for Fig.~3(b) in the main text are $\gamma=16\,$meV, $n_{\text{sat}} =5\,$W/cm$^2$, $\alpha=25\,$(W/cm$^2)^{1/2}$, $E_{SLR}=2.162\,$eV, and $\Omega_{\text{XS},0}=0.123\,$eV. In addition to the PEP emission predicted by \autoref{Iout}, we add a term linear in input power for background emission from uncoupled R6G molecules (derived from the emission spectrum outside the array). Note that $E_{SLR}$ and $\Omega_{\text{XS}}$ are slightly different from the values used to reproduce the extinction measurements, due to the differences between emission and extinction described in the main text.

%

\end{document}